\documentclass[11pt,a4paper]{article}
\usepackage[margin=2.5cm]{geometry}
\usepackage[T1]{fontenc}
\usepackage{lmodern}
\usepackage{microtype}
\usepackage{graphicx}
\usepackage{cite}
\usepackage{url}
\usepackage{xcolor}
\usepackage{amsmath, amssymb}
\usepackage{booktabs}
\usepackage{algorithm}
\usepackage{algpseudocode}
\usepackage[hidelinks]{hyperref}
\usepackage{caption}
\makeatletter
\def\bstctlcite#1{\@bsphack\@for\@citeb:=#1\do{\edef\@citeb{\expandafter\@iden\@citeb}%
    \if@filesw\immediate\write\@auxout{\string\citation{\@citeb}}\fi}\@esphack}
\makeatother

\title{Adaptive Land Units from Earth Embeddings}
\author{Albert Alonso\textsuperscript{1}, Jamal Jokar Arsanjani\textsuperscript{2}, and Andres R. Masegosa\textsuperscript{1}\\[0.8em]
\small \textsuperscript{1}Department of Computer Science, Aalborg University, Copenhagen, Denmark\\
\small \textsuperscript{2}Department of Sustainability and Planning, Aalborg University, Copenhagen, Denmark\\[0.4em]
\small \texttt{aadlf@cs.aau.dk}, \texttt{jja@plan.aau.dk}, \texttt{arma@cs.aau.dk}}
\date{}

\begin{document}
\bstctlcite{IEEEexample:BSTcontrol}
\maketitle

\begin{abstract}
    Modeling land use starts with a choice of spatial units, typically a regular grid of pixels whose resolution is set by the spatial detail the process requires and, often more critically, by the computational budget.
    With Earth observation data more abundant than ever, we ask whether good land units can instead emerge from the data itself.
    To do that, we cluster AlphaEarth embeddings with a parallel Bayesian hierarchical clustering into \textit{adaptive land units}, large where the land is uniform and small where it is highly heterogeneous.
    Using Denmark as a case study, we turn 402M land pixels into 1.27M newly defined units, about 320 times fewer, without any labels (i.e., unsupervised), making it possible to fit the whole country on a single consumer-grade GPU.
    Importantly, these units describe the land well, as a regular grid needs about 5 times as many cells to match them against land-use labels, and about 4 times as many against the declared crop of every field.
    Compared to clustering the satellite images themselves, the embeddings need half the units for the same purity.
    But more notably, they work as modeling units.
    Land-use maps learned on them are within a few points of accuracy of pixel models at a small fraction of the cost, and they make whole-country land-use estimates with reliable uncertainty possible in under a minute, where the same model on pixels does not even fit in device memory.
\end{abstract}

\section{Introduction}

The alarming rate of climate change and of nature loss is increasing the pressure on countries to rethink how their land is managed.
Land use already accounts for about a quarter of greenhouse-gas emissions, is the largest driver of biodiversity loss, and decides the security of food and water~\cite{ipcc_land_2019,ipbes_global_2019}.
Deciding what to farm, protect or restore needs models of the landscape that are detailed, cover whole countries, and are honest about their uncertainty, and every such model starts from how the land is represented.

There are two fundamental ways to represent land data in geography.
The first is the vector form, where points, lines and polygons describe what is on the ground.
It is common for registries, such as parcels, buildings and roads, and for interactive maps.
The second is the raster form, where the land is cut into a regular grid of pixels.
Rasters offer a universal way to combine data from different sources, since every layer aligns on the same grid, and they are the natural format of satellite imagery.
As a result, most modeling of land use and land cover (LULC) is done on rasters~\cite{wang_review_2023}.

But raster pixels are not independent observations.
As Tobler stated in his first law of geography~\cite{tobler_computer_1970}, near things are more related than distant things.
At the scale of pixels, this means that neighboring pixels often belong to the same field, forest or building, and therefore carry highly redundant information~\cite{hossain_segmentation_2019}.
This motivated a third representation, object-based image analysis (OBIA), where land is described by meaningful spatial units rather than individual pixels~\cite{blaschke_object_2010}.
Objects are a natural representation, since they are closer to how land is used and managed.
Despite the idea being proposed a couple of decades ago~\cite{blaschke_whats_2001}, objects remain far less common than rasters because they are especially tricky to obtain from raw imagery.
Classical OBIA constructs them by segmenting images, a process that can depend strongly on the sensor, season and hand-tuned parameters, and on the type of land~\cite{hossain_segmentation_2019, ez-zahouani_remote_2023}.

At the same time, the amount and diversity of data available about the landscape has grown drastically in recent years.
Satellites now provide repeated observations of the Earth's surface at global scale and across multiple spatial, spectral and temporal resolutions~\cite{wu_semantic-enhanced_2025}, which creates a continuous stream of information about the land.
This, in turn, has enabled foundation models for Earth observation, which learn general-purpose representations from large remote-sensing datasets~\cite{xiao_foundation_2025}.
AlphaEarth Foundations~\cite{brown_alphaearth_2025}, for example, summarizes each 10\,m pixel of the land every year as a 64-dimensional embedding, a compact vector that integrates optical, radar, elevation and climate data.

\begin{figure*}[!t]
    \centering
    \includegraphics[width=\textwidth]{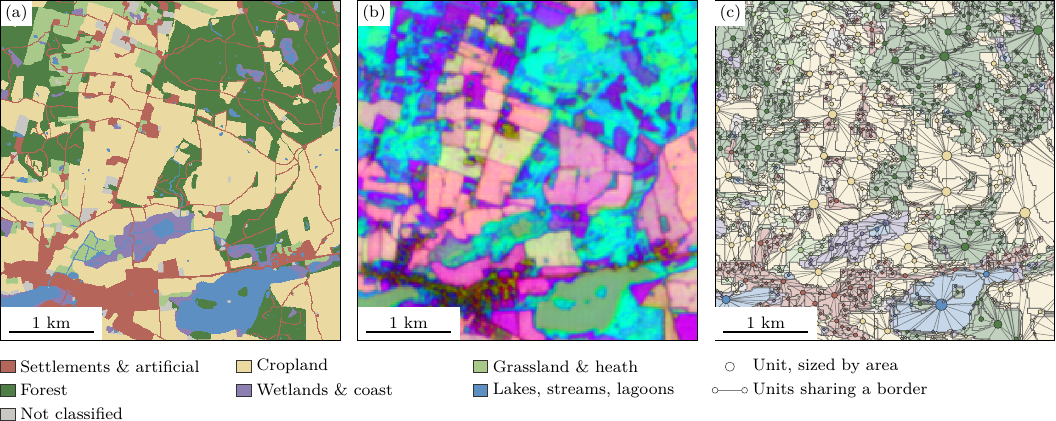}
    \caption{A 4\,km window of Denmark. (a) Basemap05 land use~\cite{levin_basemap05_2025}, with the classes merged into broad groups. (b) AlphaEarth embeddings, first three principal components as RGB. (c) The land-unit graph, with units in their majority class, nodes sized by area and edges between touching units.}
    \label{fig:svendborg}
\end{figure*}

These learned representations raise a question that is largely hidden by the raster representation.
What should the spatial unit of land be?
With a fixed pixel size, the same spatial scale is imposed everywhere, despite the strong heterogeneity of the landscape.
We know that some features require fine-grained representations, while others form coherent patterns over much larger areas~\cite{hay_comparison_2003}.
A single resolution, however, ignores this variation, and while image segmentation can let object size vary, its settings have to be tuned by hand, by trial and error, for each sensor and landscape~\cite{baatz_multiresolution_2000, hossain_segmentation_2019}.
We therefore ask whether meaningful spatial units can instead emerge directly from the learned embeddings, without having to define them manually.

Here we use Earth embeddings to cluster the land into adaptive land units, producing a compact graph of the landscape in which pixels that belong to the same object are grouped together.
We use Bayesian hierarchical clustering~\cite{heller_bayesian_2005}, constrained so that only neighboring clusters can merge, and parallelize it to run on a GPU.
By processing overlapping tiles with a ghost area around each tile~\cite{lassalle_scalable_2015}, the clustering scales to an entire country with few seam artifacts.

The resulting representation adapts its spatial resolution to the landscape.
Units become large where the land is uniform and small where it changes, rather than having a fixed imposed size (Fig.~\ref{fig:svendborg}), which makes the representation compact enough for a whole country to fit on a single consumer-grade GPU.
We show that, at the same number of cells and without using any labels, these units describe the land better than a regular grid or than units cut from the satellite image itself, and that they work as modeling units, both for mapping land use and for spatial models of a whole country that do not fit in memory on pixels.

\section{Method}
\subsection{Data and study area}
We use the 2024 AlphaEarth Foundations embeddings~\cite{brown_alphaearth_2025}, as they are global, annual, openly available, and outperform other featurizations in their benchmarks~\cite{brown_alphaearth_2025}.
Each 10\,m pixel is a 64-dimensional vector learned from over three billion observations across nine Earth-observation sources.
We chose Denmark as our case study because its land is mapped in unusual detail, and use its part in UTM zone 32N (Bornholm excluded), with 402M land pixels.
To evaluate the units, we use Basemap05~\cite{levin_basemap05_2025} for 2024 at its 13 level-1 and 29 level-2 land-use classes, of which 11 and 25 cover land, and the yearly field map of the Danish Agricultural Agency~\cite{lfst_marker_2024}, with the declared crop of every field in 14 crop groups.

\subsection{Parallel Bayesian hierarchical clustering}
Each land pixel is a node of the planar 4-neighbor pixel graph and carries its 64-dimensional embedding.
We merge them into adaptive land units using Bayesian hierarchical clustering (BHC)~\cite{heller_bayesian_2005}, restricted so that only clusters sharing an edge of the graph can merge, that is, every merged unit is a connected region of the original pixel map.
For each candidate merge, BHC compares the probability that the pixels of both clusters come from a single distribution with that of the separate clusters they already form, and merges when the posterior probability of the merge exceeds one half ($r > 0.5$).
More precisely, each unit is modeled as a Gaussian around its mean embedding, and a single parameter $s$, the within-unit variance relative to the national variance, sets how homogeneous a unit ends up being (Supplementary Section~S1).

\begin{figure}[!b]
    \centering
    \includegraphics{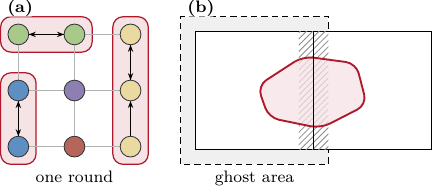}
    \caption{(a) One round of PBHC. Each cluster points to the neighbor with the most probable merge, and the groups linked by pointers (shaded) merge at once. (b) Tiling. A tile is clustered with a ghost area, units are dissolved along the seam (hatched), and a final round merges across it.}
    \label{fig:method}
\end{figure}

An important limitation of BHC is that it merges one pair at a time, which takes $n - 1$ sequential steps for $n$ pixels, that is around a million for a single 10\,km tile.
But because merges are restricted to graph neighbors, merges in different parts of the map do not interact, and we propose a parallel alternative, parallel BHC (PBHC), in which all clusters can merge at the same time.
In each round, every cluster points to the neighbor with the most probable merge, provided $r > 0.5$, and groups of clusters linked by these pointers merge in a single step (Bor\r{u}vka-style, Fig.~\ref{fig:method}a), using the BHC recursion extended to several children as in Bayesian rose trees~\cite{blundell_bayesian_2010} (Supplementary Algorithm~S1).
The rounds stop when no such merge remains, and a 10\,km tile takes 8--16 rounds and 1--3\,s on one GPU instead of a million sequential merges.

\subsection{Tiling}
The embeddings of Denmark, 402M land pixels over about 40{,}000\,km$^2$, take around 100\,GB and do not fit in the memory of a single GPU.
We therefore cluster the country in 10\,km tiles.
Clustered independently, tiles cut units at their borders, and an edge on a tile seam is then 12 times more likely than any other edge to separate two units.
Since the tile size is arbitrary, a unit should not depend on whether it lies at the border of a tile or inside it.
To fix this, each tile is clustered with a 250\,m ghost area on every side and only its core is kept~\cite{lassalle_scalable_2015} (Fig.~\ref{fig:method}b).
Units cut by a tile border are dissolved back into pixels near that border, and a final round over the stitched tiles clusters these seam zones with context from both sides, which makes the seams almost disappear (Supplementary Section~S2).
The whole country is clustered in less than an hour on one consumer GPU.

\section{Experiments}
In this section, we first evaluate how well the units describe the land.
A good unit holds a single kind of land, such as one field, one forest or one lake, rather than a mix of them.
We measure this by the purity of a partition against a reference map, here Basemap05~\cite{levin_basemap05_2025} or the declared crop of every field.
Each unit is given the class that covers most of its pixels, and the purity is the share of labeled pixels that fall in a unit of their own class.
It is therefore also the highest accuracy that any map built on these units could reach when evaluated against those labels.
Impurity $\varepsilon$ is one minus purity and is zero when each
unit contains a single pixel. Splitting units cannot increase
impurity, so we compare partitions at the same unit density $\rho$,
the number of units per km$^2$ of land. At a given density, lower
impurity indicates better agreement with the reference labels.

First, we compare the units with other partition algorithms, from regular grids to the superpixel segmenters of object-based analysis (Section~\ref{sec:partition}).
We also cluster the pixels of the satellite image directly instead of the embeddings, to measure which is the new information the embeddings add (Section~\ref{sec:input}).
Then, we evaluate how they perform as modeling units, first for mapping land use over Denmark from a few labeled points (Section~\ref{sec:cost}), and then for estimating the share of cropland in every 10\,km cell with its uncertainty, from a model of the whole country that does not fit in memory on pixels (Section~\ref{sec:buys}).

\subsection{Units that follow the landscape}
\label{sec:partition}
\begin{figure}[!t]
    \centering
    \includegraphics{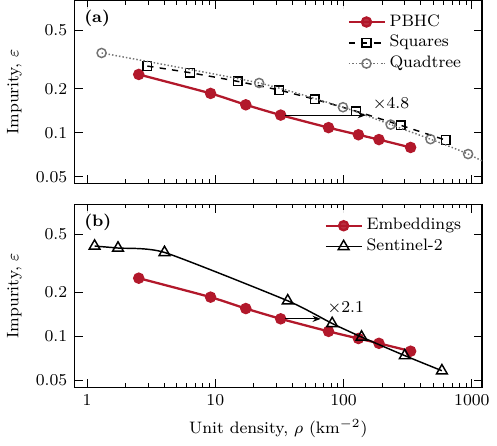}
    \caption{Impurity $\varepsilon$ against unit density $\rho$, over 10 random 20\,km tiles. Arrows show how many more units are needed to match PBHC. (a) Partitions of the embeddings (squares of 40 to 600\,m). (b) PBHC on the embeddings and on the Sentinel-2 image.}
    \label{fig:rate-distortion}
\end{figure}

\begin{table}[!t]
    \caption{
    Comparison between partitions (10 tiles of 20\,km; squares of 180\,m).
    Variance is the share of the embedding variance lost, $\varepsilon_1$ and $\varepsilon_2$ the impurity at Basemap05 levels 1 and 2, and $\varepsilon_\mathrm{c}$ the crop impurity against the declared crop of every field.
    The reported time is per tile}
    \label{tab:segmenters}
    \centering
    \footnotesize
    \setlength{\tabcolsep}{3pt}
    \begin{tabular}{@{}lrrrrrr@{}}
        \toprule
        Clustering & $\rho$ (km$^{-2}$) & Variance\,$\downarrow$ & $\varepsilon_1$\,$\downarrow$ & $\varepsilon_2$\,$\downarrow$ & $\varepsilon_\mathrm{c}$\,$\downarrow$ & Time (s)\,$\downarrow$ \\
        \midrule
        PBHC (GPU) & 32.4 & 0.20 & 0.132 & 0.175 & 0.094 & 9.6 \\
        SLIC~\cite{achanta_slic_2012} & 25.1 & 0.21 & 0.148 & 0.193 & -- & 94.5 \\
        Felzenszwalb~\cite{felzenszwalb_efficient_2004} & 22.0 & 0.48 & 0.252 & 0.307 & -- & 27.9 \\
        Quadtree~\cite{samet_quadtree_1984} & 22.0 & 0.46 & 0.219 & 0.261 & -- & 6.2 \\
        Squares & 31.6 & 0.32 & 0.196 & 0.240 & 0.178 & 0.1 \\
        PBHC, Sentinel-2 & 36.7 & -- & 0.175 & 0.234 & 0.133 & 5.4 \\
        Ward~(\emph{Reference})~\cite{ward_hierarchical_1963} & 32.4 & 0.16 & 0.128 & 0.170 & -- & 1275 \\
        \bottomrule
    \end{tabular}
\end{table}

On 10 random tiles of 20\,km, we compare PBHC, at the same number of units, with the partitions that could be used instead, namely grids (regular squares and an adaptive quadtree), superpixel segmenters as used in object-based analysis (SLIC~\cite{achanta_slic_2012} and Felzenszwalb's graph segmentation~\cite{felzenszwalb_efficient_2004}), and Ward's agglomerative clustering~\cite{ward_hierarchical_1963}, the classical alternative to a Bayesian criterion.
PBHC cuts the land into 32 units per km$^2$ with an impurity of 0.13 (Fig.~\ref{fig:rate-distortion}a).
To be as pure, regular squares need 4.8 times as many units (2.3--6.4 across tiles).
Over the whole country, the 1.27M units reach a purity of 0.881 on held-out regions, against 0.825 for the same number of squares (cells of 180\,m).
A quadtree, which splits squares wherever the embeddings vary, needs 4.5 times as many units.
Adapting the size of a cell is therefore not enough, and its boundaries have to follow the land.
The advantage holds against the declared crop of every field.
Squares need 3.7 times as many cells to match the units, and unit boundaries are almost three times as likely as square boundaries to lie on a field boundary.

Ward's agglomerative clustering on the same graph, the classical criterion for compact clusters, finds slightly purer partitions (Table~\ref{tab:segmenters}).
Nevertheless, it must be told how many units to make, has no parallel or tiled form, and takes 130 times longer per tile, while PBHC reaches nearly the same purity, decides its own number of units, and scales to the country.
Note that Ward's criterion is also, apart from a shape term, essentially the merge rule of the multiresolution segmentation in eCognition~\cite{baatz_multiresolution_2000}.
Of the superpixel segmenters, SLIC on the embeddings comes closest, but it takes ten times longer per tile and needs up to 100\,GB of memory, which kept it from running above 65 units per km$^2$.
Felzenszwalb's graph segmentation collapses at coarse scales.

\subsection{The benefits of embeddings over satellite images}
\label{sec:input}
Object-based analysis works on the image itself.
To see whether clustering the embeddings instead provides any significant benefit, we also run the same PBHC on the Sentinel-2 image~\cite{drusch_sentinel2_2012} (quarterly medians of the blue, green, red and near-infrared bands, standardized).
Our results show that the image needs 2.1 times as many units to be as pure as on the embeddings (Fig.~\ref{fig:rate-distortion}b), and 1.7 times as many against the crops, and it collapses for $s \geq 1$ into a few large units with an impurity near 0.4.
Above about 200 units per km$^2$ the raw image is as pure or purer, since at 10\,m it is sharper than the embeddings.
The embeddings therefore matter most at the coarse scales where the compression happens.

\subsection{Mapping land use from sparse labels}
\label{sec:cost}

A first use of the units is mapping land use from a few labeled points, the usual situation when labels are expensive.
We map level-1 land use over Denmark from $K$ labeled points and score it on held-out regions, classifying units and squares from their mean embedding and area, and pixels from their own embedding or a small window around them (Supplementary Section~S4).
Units beat squares by 4--5 percentage points from $K = 10^3$ on (Table~\ref{tab:mapping}).
Against pixels, units tie at $K = 10^2$, fall between the two pixel models at $K = 10^3$, and stay within 4 points with all labels, as their accuracy saturates near 0.83, below their purity of 0.88.
In exchange, the whole country is one graph on the GPU.
Predicting all of Denmark takes 0.2\,s in one forward pass, 270--700 times faster than the pixel models streamed from a local copy of the raster, and with 80 times less data copied to the GPU.

\begin{table}[!bt]
    \caption{Level-1 accuracy from $K$ labeled points (3 seeds; All, one run) and cost of predicting all of Denmark.}
    \label{tab:mapping}
    \centering
    \footnotesize
    \setlength{\tabcolsep}{4pt}
    \begin{tabular}{@{}lrrrrr@{}}
        \toprule
        & \multicolumn{3}{c}{Accuracy\,$\uparrow$} & \multicolumn{2}{c}{National inference\,$\downarrow$} \\
        \cmidrule(lr){2-4} \cmidrule(l){5-6}
        Model & $K = 10^2$ & $10^3$ & All & Time (s) & To GPU (GB) \\
        \midrule
        Units, MLP & 0.750 & 0.799 & 0.828 & 0.2 & 0.47 \\
        Squares, MLP & 0.718 & 0.755 & 0.782 & 0.2 & 0.43 \\
        Pixels, MLP & 0.746 & 0.813 & 0.855 & 145 & 38 \\
        Pixels, CNN & 0.746 & 0.789 & 0.867 & 54 & 38 \\
        \bottomrule
    \end{tabular}
\end{table}

\subsection{Whole-country estimates with uncertainty}
\label{sec:buys}
\begin{figure}[!t]
    \centering
    \includegraphics{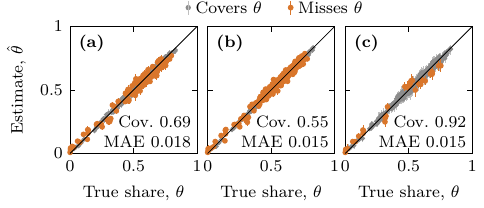}
    \caption{Cropland share of the 406 cells of 10\,km from $10^4$ survey points (one seed), estimate $\hat\theta$ with its 90\,\% interval against the true share $\theta$, on (a) squares with the spatial field, (b) adaptive land units without
the spatial field, and (c) adaptive land units with the spatial field. Coverage and mean absolute error in each corner.}
    \label{fig:cells}
\end{figure}

\begin{figure}[!t]
    \centering
    \includegraphics{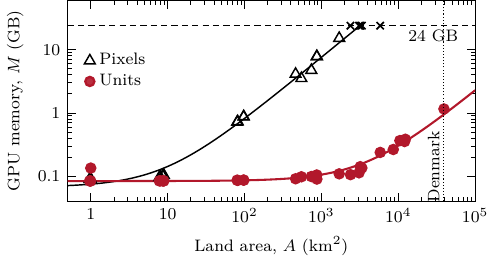}
    \caption{GPU memory of the spatial model for square windows of 1 to 120\,km at 5 locations and for all of Denmark. Lines are fits $M = M_0 + k A$, and $\times$ marks pixel windows that exceed the 24\,GB of the GPU.}
    \label{fig:scaling}
\end{figure}

Other than for storage efficiency, a big incentive to create accurate yet compact representations that fits on one GPU is that they make joint models of the whole country possible.
To illustrate that, we estimate the cropland share of every 10\,km cell, alongside its uncertainty, from a survey of up to 10{,}000 points over the labeled land (Supplementary Section~S5).
Let each unit $u$ have a probability $p_u$ of being cropland, modeled as
\begin{equation}
    \operatorname{logit} p_u = \mathbf{x}_u^\top \boldsymbol{\beta} + g_u ,
\end{equation}
where $\boldsymbol{\beta}$ are the regression coefficients on the unit features $\mathbf{x}_u$, and the estimated share of a cell $\hat \theta$ is the average of $p_u$ over its pixels.
While the regression alone gets most units roughly right, nearby units tend to be wrong in the same direction.
The spatial field $g_u$ corrects for this with a term that couples connected units in the graph.
Because it links every unit to its neighbors, and through them to the rest of the country, it has to be fitted to all of Denmark at once, which means inverting a sparse matrix with one row per unit.
On regular squares, even with the field, the estimates are less accurate and their intervals too narrow (Fig.~\ref{fig:cells}a).
On units without the field, the estimates are as accurate as with it but overconfident, and almost half of the 90\,\% intervals miss the true share (Fig.~\ref{fig:cells}b).
Finally, when accounting for the field in the units, the intervals are calibrated, covering the true share in 92\,\% of the cells (Fig.~\ref{fig:cells}c).
On pixels the model runs out of memory in most 60\,km windows even on a 24\,GB GPU.
For window sizes that still fit in memory, it is up to 150 times slower, while on units, the entirety of Denmark takes 1.2\,GB and 45\,s to run (Fig.~\ref{fig:scaling}).

\section{Discussion}
Adaptive land units compress the landscape while retaining its underlying structure.
At the same number of cells they describe land use and crops better than a grid, lose only a few points of mapping accuracy against pixels, and bring a whole country onto one GPU.
While we have shown this for two illustrative cases, its use is more general.
Many land-use models are iterative, simulating change year after year, calibrating parameters or comparing scenarios, and each pass over a compact national graph takes seconds instead of a pass over hundreds of millions of pixels.

Much of this comes from clustering the embeddings rather than the image.
Since the embeddings already describe what is on the ground, the clustering only has to find where it the changes are significant, and we see that compared to the raw satellite images, it needs twice less units to do so.
The resulting graph neighbors also signal stronger information then when unclustered.
On a grid, two cells are neighbors because of where the raster happens to fall, while two units are neighbors because two parts of the landscape meet there, which is why the same spatial field gives calibrated intervals on units but not on squares.
Pixels remain the better choice when labels are plentiful.
In most applications, however, much of the land is uniform over large areas, a field, a forest or a lake, and describing it pixel by pixel spends memory and computation where nothing changes.
Units are therefore most useful when labels are sparse, models need to be fast, or the whole country has to be modeled at once.

Like any compression, the units trade some fine detail for compactness.
They are also built one year at a time, and temporal applications may need a partition tuned for consistency over time (Supplementary Section~S3).
Finally, purity and accuracy measure agreement with Basemap05, which carries errors of its own.

\section{Conclusion}
We asked whether good land units can emerge from the data instead of being fixed in advance.
Clustering Earth embeddings with a parallel Bayesian hierarchical clustering, we turned the 402M land pixels of Denmark into 1.27M adaptive land units without any labels.
They describe land use and crops better than a grid with the same number of cells, and they make whole-country models with reliable uncertainty possible on a single GPU.
As Earth embeddings now cover the planet every year, adaptive land units open a new way to represent land, in which the spatial unit of every model is learned from the data rather than imposed by a grid.
Code and data are openly available~\cite{alonso_zenodo_2026}.

\section*{Acknowledgments}
This work was supported by a research grant (VIL78584) from Villum Fonden (project DK-Future, Villum Synergy Programme).

\bibliographystyle{IEEEtran}
\bibliography{references}

\begin{thebibliography}{10}
\providecommand{\url}[1]{#1}
\csname url@samestyle\endcsname
\providecommand{\newblock}{\relax}
\providecommand{\bibinfo}[2]{#2}
\providecommand{\BIBentrySTDinterwordspacing}{\spaceskip=0pt\relax}
\providecommand{\BIBentryALTinterwordstretchfactor}{4}
\providecommand{\BIBentryALTinterwordspacing}{\spaceskip=\fontdimen2\font plus
\BIBentryALTinterwordstretchfactor\fontdimen3\font minus
  \fontdimen4\font\relax}
\providecommand{\BIBforeignlanguage}[2]{{%
\expandafter\ifx\csname l@#1\endcsname\relax
\typeout{** WARNING: IEEEtran.bst: No hyphenation pattern has been}%
\typeout{** loaded for the language `#1'. Using the pattern for}%
\typeout{** the default language instead.}%
\else
\language=\csname l@#1\endcsname
\fi
#2}}
\providecommand{\BIBdecl}{\relax}
\BIBdecl

\bibitem{ipcc_land_2019}
{IPCC}, \emph{Climate Change and Land: {IPCC} Special Report on Climate Change,
  Desertification, Land Degradation, Sustainable Land Management, Food
  Security, and Greenhouse Gas Fluxes in Terrestrial Ecosystems}, P.~R. Shukla,
  J.~Skea, E.~Calvo~Buendia, V.~Masson-Delmotte \emph{et~al.}, Eds.\hskip 1em
  plus 0.5em minus 0.4em\relax Cambridge University Press, 2019.

\bibitem{ipbes_global_2019}
{IPBES}, \emph{Global Assessment Report on Biodiversity and Ecosystem Services
  of the Intergovernmental Science-Policy Platform on Biodiversity and
  Ecosystem Services}, E.~S. Brondízio, J.~Settele, S.~D{\'\i}az, and H.~T.
  Ngo, Eds.\hskip 1em plus 0.5em minus 0.4em\relax Bonn, Germany: IPBES
  Secretariat, 2019.

\bibitem{wang_review_2023}
Y.~Wang, Y.~Sun, X.~Cao, Y.~Wang, W.~Zhang, and X.~Cheng, ``A review of
  regional and global scale land use/land cover ({LULC}) mapping products
  generated from satellite remote sensing,'' \emph{ISPRS Journal of
  Photogrammetry and Remote Sensing}, vol. 206, pp. 311--334, Dec. 2023.

\bibitem{tobler_computer_1970}
W.~R. Tobler, ``A computer movie simulating urban growth in the {Detroit}
  region,'' \emph{Economic Geography}, vol.~46, pp. 234--240, Jun. 1970.

\bibitem{hossain_segmentation_2019}
M.~D. Hossain and D.~Chen, ``Segmentation for object-based image analysis
  ({OBIA}): {A} review of algorithms and challenges from remote sensing
  perspective,'' \emph{ISPRS Journal of Photogrammetry and Remote Sensing},
  vol. 150, pp. 115--134, Apr. 2019.

\bibitem{blaschke_object_2010}
T.~Blaschke, ``Object based image analysis for remote sensing,'' \emph{ISPRS
  Journal of Photogrammetry and Remote Sensing}, vol.~65, no.~1, pp. 2--16,
  Jan. 2010.

\bibitem{blaschke_whats_2001}
T.~Blaschke and J.~Strobl, ``What's wrong with pixels? {Some} recent
  developments interfacing remote sensing and {GIS},'' \emph{{GIS} --
  Zeitschrift f{\"u}r Geoinformationssysteme}, no.~6, pp. 12--17, 2001.

\bibitem{ez-zahouani_remote_2023}
B.~Ez-zahouani, A.~Teodoro, O.~El~Kharki, L.~Jianhua, I.~Kotaridis, X.~Yuan,
  and L.~Ma, ``Remote sensing imagery segmentation in object-based analysis:
  {A} review of methods, optimization, and quality evaluation over the past 20
  years,'' \emph{Remote Sensing Applications: Society and Environment},
  vol.~32, p. 101031, Nov. 2023.

\bibitem{wu_semantic-enhanced_2025}
K.~Wu, Y.~Zhang, L.~Ru, B.~Dang, J.~Lao, L.~Yu, J.~Luo, Z.~Zhu, Y.~Sun,
  J.~Zhang, Q.~Zhu, J.~Wang, M.~Yang, J.~Chen, Y.~Zhang, and Y.~Li,
  ``\BIBforeignlanguage{en}{A semantic-enhanced multi-modal remote sensing
  foundation model for {Earth} observation},''
  \emph{\BIBforeignlanguage{en}{Nature Machine Intelligence}}, vol.~7, no.~8,
  pp. 1235--1249, Aug. 2025.

\bibitem{xiao_foundation_2025}
A.~Xiao, W.~Xuan, J.~Wang, J.~Huang, D.~Tao, S.~Lu, and N.~Yokoya, ``Foundation
  models for remote sensing and {Earth} observation: {A} survey,'' \emph{IEEE
  Geoscience and Remote Sensing Magazine}, vol.~13, no.~4, pp. 297--324, Dec.
  2025.

\bibitem{brown_alphaearth_2025}
C.~F. Brown, M.~R. Kazmierski, V.~J. Pasquarella, W.~J. Rucklidge,
  M.~Samsikova, C.~Zhang, E.~Shelhamer, E.~Lahera, O.~Wiles, S.~Ilyushchenko,
  N.~Gorelick, L.~L. Zhang, S.~Alj, E.~Schechter, S.~Askay, O.~Guinan,
  R.~Moore, A.~Boukouvalas, and P.~Kohli, ``{AlphaEarth} {Foundations}: {An}
  embedding field model for accurate and efficient global mapping from sparse
  label data,'' Sep. 2025, arXiv:2507.22291 [cs.CV].

\bibitem{levin_basemap05_2025}
G.~Levin, ``{BASEMAP05}: Documentation of the data and method for elaboration
  of a land use and land cover map for {Denmark},'' Aarhus University, DCE --
  Danish Centre for Environment and Energy, Technical Report 363, Dec. 2025.

\bibitem{hay_comparison_2003}
G.~J. Hay, T.~Blaschke, D.~J. Marceau, and A.~Bouchard, ``A comparison of three
  image-object methods for the multiscale analysis of landscape structure,''
  \emph{ISPRS Journal of Photogrammetry and Remote Sensing}, vol.~57, no.~5,
  pp. 327--345, Apr. 2003.

\bibitem{baatz_multiresolution_2000}
M.~Baatz and A.~Sch{\"a}pe, ``Multiresolution segmentation: an optimization
  approach for high quality multi-scale image segmentation,'' in
  \emph{Angewandte {Geographische} {Informationsverarbeitung} {XII}},
  J.~Strobl, T.~Blaschke, and G.~Griesebner, Eds.\hskip 1em plus 0.5em minus
  0.4em\relax Heidelberg: Wichmann, 2000, pp. 12--23.

\bibitem{heller_bayesian_2005}
K.~A. Heller and Z.~Ghahramani, ``Bayesian hierarchical clustering,'' in
  \emph{Proceedings of the 22nd {International} {Conference} on {Machine}
  {Learning}}, ser. {ICML} '05.\hskip 1em plus 0.5em minus 0.4em\relax New
  York, NY, USA: Association for Computing Machinery, Aug. 2005, pp. 297--304.

\bibitem{lassalle_scalable_2015}
P.~Lassalle, J.~Inglada, J.~Michel, M.~Grizonnet, and J.~Malik, ``A scalable
  tile-based framework for region-merging segmentation,'' \emph{IEEE
  Transactions on Geoscience and Remote Sensing}, vol.~53, no.~10, pp.
  5473--5485, Oct. 2015.

\bibitem{lfst_marker_2024}
{Danish Agricultural Agency (Landbrugsstyrelsen)}, ``Marker: field map with
  declared crops, 2017--2025,'' Open data, geodata.fvm.dk, 2024.

\bibitem{blundell_bayesian_2010}
C.~Blundell, Y.~W. Teh, and K.~A. Heller, ``Bayesian rose trees,'' in
  \emph{Proceedings of the 26th Conference on Uncertainty in Artificial
  Intelligence}, 2010, pp. 65--72.

\bibitem{achanta_slic_2012}
R.~Achanta, A.~Shaji, K.~Smith, A.~Lucchi, P.~Fua, and S.~S{\"u}sstrunk,
  ``{SLIC} superpixels compared to state-of-the-art superpixel methods,''
  \emph{IEEE Transactions on Pattern Analysis and Machine Intelligence},
  vol.~34, no.~11, pp. 2274--2282, 2012.

\bibitem{felzenszwalb_efficient_2004}
P.~F. Felzenszwalb and D.~P. Huttenlocher, ``Efficient graph-based image
  segmentation,'' \emph{International Journal of Computer Vision}, vol.~59,
  no.~2, pp. 167--181, 2004.

\bibitem{samet_quadtree_1984}
H.~Samet, ``The quadtree and related hierarchical data structures,'' \emph{ACM
  Computing Surveys}, vol.~16, no.~2, pp. 187--260, 1984.

\bibitem{ward_hierarchical_1963}
J.~H. Ward, ``Hierarchical grouping to optimize an objective function,''
  \emph{Journal of the American Statistical Association}, vol.~58, no. 301, pp.
  236--244, 1963.

\bibitem{drusch_sentinel2_2012}
M.~Drusch, U.~Del~Bello, S.~Carlier, O.~Colin, V.~Fernandez, F.~Gascon,
  B.~Hoersch, C.~Isola, P.~Laberinti, P.~Martimort, A.~Meygret, F.~Spoto,
  O.~Sy, F.~Marchese, and P.~Bargellini, ``Sentinel-2: {ESA}'s optical
  high-resolution mission for {GMES} operational services,'' \emph{Remote
  Sensing of Environment}, vol. 120, pp. 25--36, 2012.

\bibitem{alonso_zenodo_2026}
A.~Alonso, J.~J. Arsanjani, and A.~R. Masegosa, ``Code and data for {Adaptive
  Land Units from Earth Embeddings},'' Zenodo,
  \url{https://doi.org/10.5281/zenodo.23836328}, 2026.

\bibitem{bradbury_jax_2018}
J.~Bradbury, R.~Frostig, P.~Hawkins, M.~J. Johnson, C.~Leary, D.~Maclaurin,
  G.~Necula, A.~Paszke, J.~VanderPlas, S.~Wanderman-Milne, and Q.~Zhang,
  ``{JAX}: composable transformations of {Python}+{NumPy} programs,'' 2018.

\bibitem{papandreou_gaussian_2010}
G.~Papandreou and A.~L. Yuille, ``Gaussian sampling by local perturbations,''
  in \emph{Advances in {Neural} {Information} {Processing} {Systems}}, vol.~23,
  2010, pp. 1858--1866.

\end{thebibliography}

\clearpage
\appendix
\section*{Supplementary Material}
\setcounter{section}{0}
\setcounter{table}{0}
\setcounter{figure}{0}
\setcounter{equation}{0}
\setcounter{algorithm}{0}
\renewcommand{\thesection}{S\arabic{section}}
\renewcommand{\thesubsection}{\thesection.\arabic{subsection}}
\renewcommand{\thetable}{S\arabic{table}}
\renewcommand{\thefigure}{S\arabic{figure}}
\renewcommand{\theequation}{S\arabic{equation}}
\renewcommand{\thealgorithm}{S\arabic{algorithm}}

\section{Parallel Bayesian hierarchical clustering}
\label{sec:s-pbhc}
Our parallel version of Bayesian hierarchical clustering (BHC) stays close to the original formulation of Heller and Ghahramani~\cite{heller_bayesian_2005}.
It differs in two ways, as only clusters that share an edge of a planar graph can merge, and many merges happen at once instead of one pair at a time.

\subsection{Algorithm}
Let $\mathcal{D} = \{\mathbf{x}^{(1)}, \dots, \mathbf{x}^{(n)}\}$ be the embeddings of the $n$ land pixels, $\mathbf{x}^{(i)} \in \mathbb{R}^{64}$, and let $G = (V, E)$ be the 4-neighbor graph of the pixel grid.
At any time the pixels are partitioned into clusters, and we write $\mathcal{D}_k$ for the data in cluster $k$, $n_k = |\mathcal{D}_k|$, and $T_k$ for the tree that built it.
Merges are restricted to clusters that share an edge of $G$, and every cluster is therefore a spatially connected region.

As in BHC, merging a set of clusters $\mathcal{C}$ into $\mathcal{D}_k = \bigcup_{c \in \mathcal{C}} \mathcal{D}_c$ compares two hypotheses.
Under $\mathcal{H}_1^k$ all data in $\mathcal{D}_k$ come from one distribution $p(\mathbf{x} \mid \theta)$ with a single unknown $\theta$, and
\begin{equation}
    p(\mathcal{D}_k \mid \mathcal{H}_1^k) = \int p(\theta \mid \beta) \prod_{\mathbf{x}^{(i)} \in \mathcal{D}_k} p(\mathbf{x}^{(i)} \mid \theta) \, d\theta .
\end{equation}
The alternative is that the data are partitioned as in the children's trees, with probability $\prod_{c \in \mathcal{C}} p(\mathcal{D}_c \mid T_c)$.
Weighting the two by the Dirichlet process prior gives the marginal likelihood of the merged tree,
\begin{equation}
    p(\mathcal{D}_k \mid T_k) = \pi_k \, p(\mathcal{D}_k \mid \mathcal{H}_1^k) + (1 - \pi_k) \prod_{c \in \mathcal{C}} p(\mathcal{D}_c \mid T_c),
    \label{eq:tree}
\end{equation}
where $\pi_k$ follows the BHC recursion, extended from two children to several as in Bayesian rose trees~\cite{blundell_bayesian_2010},
\begin{equation}
    d_k = \alpha \, \Gamma(n_k) + \prod_{c \in \mathcal{C}} d_c, \qquad
    \pi_k = \frac{\alpha \, \Gamma(n_k)}{d_k}, \qquad
    1 - \pi_k = \frac{\prod_{c \in \mathcal{C}} d_c}{d_k},
    \label{eq:prior}
\end{equation}
with $d_i = \alpha$ and $\pi_i = 1$ for a single pixel.
We compute $1 - \pi_k$ from the ratio on the right and not as one minus $\pi_k$, because the subtraction loses all precision once $\pi_k$ rounds to one in floating point.
The posterior probability of the merge is $r_k = \pi_k \, p(\mathcal{D}_k \mid \mathcal{H}_1^k) / p(\mathcal{D}_k \mid T_k)$, and, as in BHC, a merge is accepted only if $r_k > 0.5$.

Sequential BHC merges the single best pair of the whole dataset at each step, which takes $n - 1$ steps and compares every pair of clusters.
On a graph every merge is local and clusters far apart do not compete, which lets many merges happen in the same round.
In each round, every edge $(i, j)$ between two different clusters is scored by the log-odds of merging the pair, $\rho_{ij} = \log r_{ij} - \log(1 - r_{ij})$, which is positive exactly when $r_{ij} > 0.5$.
Each cluster points to its best-scoring neighbor if that score is positive, and to itself otherwise, and the pointers form trees in which a pair pointing at each other keeps the smaller index as root.
All clusters in a tree merge in one step, Bor\r{u}vka-style, with Eqs.~\eqref{eq:tree}--\eqref{eq:prior} applied over all its children.
The rounds stop when no edge has $\rho_{ij} > 0$, and every pair of neighboring units in the output has $r \le 0.5$.
A 10\,km tile (1M pixels) takes 8--16 rounds instead of a million sequential steps (Algorithm~\ref{alg:pbhc}).

\begin{algorithm}[htbp]
\caption{Parallel Bayesian hierarchical clustering on a graph}
\label{alg:pbhc}
\begin{algorithmic}
\Require data $\mathcal{D} = \{\mathbf{x}^{(1)}, \dots, \mathbf{x}^{(n)}\}$, graph $G = (V, E)$, model $p(\mathbf{x} \mid \theta)$, prior $p(\theta \mid \beta)$, concentration $\alpha$
\State make each pixel its own cluster, $\mathcal{D}_i = \{\mathbf{x}^{(i)}\}$, $d_i = \alpha$, $\pi_i = 1$, $p(\mathcal{D}_i \mid T_i) = p(\mathcal{D}_i \mid \mathcal{H}_1^i)$
\Repeat
    \State compute $\rho_{ij}$ for all edges $(i, j) \in E$ with $i \neq j$, in parallel
    \State $t(i) \gets \arg\max_j \rho_{ij}$ if $\max_j \rho_{ij} > 0$, else $t(i) \gets i$, for all clusters $i$ in parallel
    \State break 2-cycles by setting $t(i) \gets i$ when $t(t(i)) = i$ and $i < t(i)$
    \State $\mathrm{root}(i) \gets t(t(\cdots t(i)))$ \Comment{pointer jumping}
    \ForAll{roots $k$ with children $\mathcal{C}_k = \{i : \mathrm{root}(i) = k\}$, $|\mathcal{C}_k| > 1$, \textbf{in parallel}}
        \State $\mathcal{D}_k \gets \bigcup_{c \in \mathcal{C}_k} \mathcal{D}_c$
        \State compute $d_k$, $\pi_k$ (Eq.~\ref{eq:prior}) and $p(\mathcal{D}_k \mid T_k)$ (Eq.~\ref{eq:tree})
    \EndFor
    \State relabel the endpoints of every edge by their roots
\Until{no cluster merges}
\Ensure a partition of $\mathcal{D}$ into spatially connected units
\end{algorithmic}
\end{algorithm}

\subsection{Model}
We use the conjugate Gaussian model, independent across the $D = 64$ dimensions, with $x_d \mid \mu_d \sim \mathcal{N}(\mu_d, \sigma_d^2)$ and $\mu_d \sim \mathcal{N}(0, \tau_d^2)$.
Before clustering, we subtract the mean embedding of all land pixels in Denmark, which makes the prior mean of zero the average Danish land pixel.
A cluster enters only through its size $n_k$ and the sums $S_{kd} = \sum_i x^{(i)}_d$ and $Q_{kd} = \sum_i (x^{(i)}_d)^2$, with
\begin{equation}
    \log p(\mathcal{D}_k \mid \mathcal{H}_1^k) = \sum_{d=1}^{D} \left[ -\frac{n_k}{2} \log (2 \pi \sigma_d^2) - \frac{1}{2} \log \left(1 + \frac{n_k \tau_d^2}{\sigma_d^2}\right) - \frac{1}{2 \sigma_d^2} \left( Q_{kd} - \frac{\tau_d^2 S_{kd}^2}{\sigma_d^2 + n_k \tau_d^2} \right) \right].
\end{equation}
Merging clusters only adds their sufficient statistics, and a round costs a constant amount of work per edge and per cluster.
We set $\tau_d^2$ to the national variance of dimension $d$ over land and $\sigma_d^2 = s \, \tau_d^2$.
The single parameter $s$ sets the scale of the units, and a larger $s$ tolerates more variation within a unit and gives fewer, larger units.
The national mean and variance are estimated from a sample of 40 random 20\,km tiles, and we use $s = 1$ and $\alpha = 1$ throughout.

The full clustering is about 80 lines of JAX~\cite{bradbury_jax_2018} and runs on a GPU, where each round is a fixed sequence of vectorized segment operations over the edges and clusters.

\section{Tiling and seams}
\label{sec:s-tiling}
To cover the country, we cluster 10\,km tiles, each with a 250\,m ghost area on every side~\cite{lassalle_scalable_2015}, and keep only the core of each tile.
Clusters wholly inside the core are kept as they are, but a cluster cut by the core edge may have been shaped by the tile border, and we dissolve its pixels within 250\,m of that edge back into single pixels while the rest stays one piece.
A final round of the same algorithm over the stitched pieces clusters these seam zones with context from both sides.
The national build repeats the scheme one level up, on 60\,km blocks of 36 tiles, each block with its own 250\,m ghost area.

To measure what is left of the seams, we compare how often a pixel edge on a seam separates two units with how often any other edge does.
On one 60\,km block, independent tiles give a ratio of 12.3.
Stitching without a ghost area gives 6.3, and the full scheme 1.9.
Over the whole country, seam edges remain 1.4 to 2.0 times as likely as other edges to separate units, but the extra boundaries on seams make up about 0.04\,\% of all unit boundaries.

\section{Partitions}
\label{sec:s-partitions}
All comparisons use the same 10 random 20\,km tiles as the main text, with PBHC at $s = 1$ as the reference.
To be as pure as PBHC at levels 1 and 2, squares need 4.8 and 4.2 times as many units (2.3--6.4 over tiles at level~1), the quadtree 4.5 and 3.9 times (3.9--5.1), and PBHC on the Sentinel-2 image 2.1 and 2.3 times (1.7--2.4).

Ward's agglomerative clustering~\cite{ward_hierarchical_1963} runs on the same embeddings and 4-neighbor graph, on each connected component of land, with the tree cut at PBHC's number of units.
It needs 0.86 times as many units as PBHC for the same purity (0.44--1.01 over tiles), but takes 11--28\,min and 11--16\,GB per tile on the CPU, against about 10\,s on the GPU, needs the number of units as input, and has no stitching scheme for a national build.

Against the declared crops of 2024 (14 crop groups, field pixels only), PBHC is purer than both squares and the units cut from the Sentinel-2 image from about 7 to 100 units per km$^2$.
At coarser densities its large units merge neighboring fields and squares become purer, and at finer densities the units cut from the Sentinel-2 image are purer.
At the matched density, 72\,\% of the field area lies in fields that are at least 80\,\% covered by a single unit, against 3\,\% for squares, and 66\,\% of the unit boundary edges lie within one pixel of a field boundary, against 24\,\% for squares, about what chance alone gives.

We also built and scored the units separately for every year from 2017 to 2025, and their crop purity stays between 0.90 and 0.92.
Between consecutive years, however, about half of the land lies in units with no match in the other year (intersection over union below 0.5).
Units regroup no more often on fields whose crop changed than on fields that kept it (43\,\% against 48\,\% of field pixels), and the changes therefore do not follow crop rotation.

\section{Mapping setup}
\label{sec:s-mapping}
We map the 11 Basemap05 level-1 land classes of 2024 over all of Denmark from $K$ labeled points, with a model on units, on as many squares, or on pixels (Table~\ref{tab:mapping}).

We divide Denmark into 10\,km blocks and group them into 40\,km groups, which we assign to training, validation and test in proportions of 60, 20 and 20\,\% of the land.
Training points within 5\,km of a validation or test group are dropped, and a unit or square belongs to the group of its majority pixel.

For each of 3 seeds, a budget of $K$ points takes the first $K$ of $10^6$ points drawn uniformly over the training land, and all models train on the same points, a unit or square model on the units or squares that hold at least one point, labeled by their majority class.
The column All of Table~\ref{tab:mapping} is a single run with $10^7$ points, which labels every unit.

Units and the 1.26M squares of 180\,m, as many as there are units, are classified by a multilayer perceptron (MLP) on their mean embedding and the log of their area.
Pixels are classified by an MLP on their own embedding or by a convolutional network (CNN) of three $3 \times 3$ layers, which sees a $7 \times 7$ pixel window.
All models use hidden size 128, dropout 0.1, unweighted cross-entropy, and Adam with learning rate $10^{-3}$ and weight decay $10^{-4}$, with early stopping on the validation loss.
A two-layer graph neural network on the land-unit graph is as accurate as the MLP (0.825 against 0.826 at $K = 10^5$).
Accuracy is the share of test pixels predicted correctly, where a unit or square passes its prediction to all of its pixels.

Inference is timed for all of Denmark on one RTX 4090 GPU (24\,GB).
Units and squares fit on the GPU as one graph (0.47 and 0.43\,GB), while the pixel models read the raster from a local int8 copy in 580 tiles, 38\,GB in total.

\section{Spatial model}
\label{sec:s-spatial}
We estimate the cropland share of every 10\,km cell of Denmark, with a 90\,\% interval, from a survey of $K$ points (Figs.~4 and 5 ).
The cells are only used for reporting, and the model is fitted on the 1.27M adaptive land units, with survey points drawn uniformly over the labeled land for $K = 10^2$, $10^3$ and $10^4$ and 3 seeds.
Each point records whether its pixel is Cropland in Basemap05 level 1, and we count for each unit $u$ the $m_u$ points it holds and the $k_u$ of them that are cropland.

The probability $p_u$ that a pixel of unit $u$ is cropland is modeled as
\begin{equation}
    \operatorname{logit} p_u = \mathbf{x}_u^\top \boldsymbol{\beta} + g_u ,
\end{equation}
and the survey points of each unit give a binomial likelihood, $k_u \sim \mathrm{Bin}(m_u, p_u)$.
Here $\mathbf{x}_u$ holds an intercept and the standardized mean embedding and log area of the unit, as in Section~\ref{sec:s-mapping}, and $g_u$ is a spatial field shared by neighboring units.
The coefficients have the prior $\boldsymbol{\beta} \sim \mathcal{N}(0, I)$, except for a standard deviation of 10 on the intercept.
The field follows a proper conditional autoregressive (CAR) prior on the land-unit graph,
\begin{equation}
    g \sim \mathcal{N}\big(0, \sigma_g^2 Q^{-1}\big), \qquad Q = D - W + \delta (D + I),
\end{equation}
where $W_{uv}$ is the length of the border between units $u$ and $v$, divided by its mean over all bordering pairs, and $D$ is the diagonal matrix of the row sums of $W$.
$Q$ is nonzero only for the 3.5M pairs of units that share a border, and the term $\delta (D + I)$, with $\delta = 10^{-3}$, makes it invertible, where a smaller $\delta$ lets the field vary over longer distances.

We choose $\sigma_g$ from $\{0.5, 1, 2, 4\}$ by fitting on four fifths of the survey points and scoring the log-likelihood of the remaining fifth, and then refit on all points.
The posterior of $\boldsymbol{\beta}$ and $g$ is approximated by a Gaussian at its mode (Laplace approximation), which we find by Newton's method.
The Hessian of the negative log posterior is as sparse as $Q$, and each Newton step is solved by conjugate gradients, which only needs sums over the edges of the graph, and memory grows only with the number of units and edges.
We never form the posterior covariance and instead draw 100 joint samples by perturb-and-MAP~\cite{papandreou_gaussian_2010}.

Units and cells do not line up, and each unit counts toward cell $c$ by its number of labeled pixels in that cell, $n_{uc}$.
For each posterior sample, the cropland share of the cell is $\hat\theta_c = \sum_u n_{uc} \, p_u / \sum_u n_{uc}$.
The estimate is the mean of the 100 samples, and the 90\,\% interval runs from their 5th to their 95th percentile.
We score them against the true share $\theta_c$, the share of labeled pixels of the cell that are cropland in Basemap05, on the 406 cells with at least 50\,km$^2$ of labeled land.
Coverage is the share of cells whose interval contains $\theta_c$, with a target of 0.90, and the mean absolute error (MAE) compares the estimates with $\theta_c$.

Table~\ref{tab:survey} compares the model with three baselines, starting with the same model on units without the field ($g = 0$).
On squares, the field links 180\,m squares that share a side, with an independent effect per square.
The survey alone estimates each cell by the share of its points that are cropland, with a Wilson 90\,\% interval, on the cells that hold at least one point.

\begin{table}[ht]
    \caption{Cropland share of 406 cells of 10\,km from $K$ survey points, mean of 3 seeds.}
    \label{tab:survey}
    \centering
    \footnotesize
    \begin{tabular}{@{}lccc@{}}
        \toprule
        & \multicolumn{3}{c}{Coverage (target 0.90) / MAE\,$\downarrow$} \\
        \cmidrule(l){2-4}
        Model & $K = 10^2$ & $10^3$ & $10^4$ \\
        \midrule
        Units, spatial field & 0.98 / 0.059 & 0.96 / 0.025 & 0.92 / 0.016 \\
        Units, no field & 0.96 / 0.057 & 0.85 / 0.023 & 0.52 / 0.016 \\
        Squares, spatial field & 0.95 / 0.064 & 0.86 / 0.037 & 0.81 / 0.021 \\
        Survey only & 0.97 / 0.412 & 0.94 / 0.283 & 0.91 / 0.078 \\
        \bottomrule
    \end{tabular}
\end{table}

\paragraph{Memory and time}
In the windows of Fig.~5 , with $10^3$ survey points, the model on the pixel grid takes a median of 258\,s at 30\,km, the largest window where all pixel runs fit, against 3.3\,s on units.

\end{document}